\documentclass[pdflatex,sn-mathphys-num]{sn-jnl}% Math and Physical Sciences Numbered Reference Style 
\usepackage{graphicx}%
\usepackage{multirow}%
\usepackage{amsmath,amssymb,amsfonts}%
\usepackage{amsthm}%
\usepackage{mathrsfs}%
\usepackage[title]{appendix}%
\usepackage{xcolor}%
\usepackage{textcomp}%
\usepackage{manyfoot}%
\usepackage{booktabs}%
\usepackage{algorithm}%
\usepackage{algorithmicx}%
\usepackage{algpseudocode}%
\usepackage{listings}%
\usepackage{lineno}
\theoremstyle{thmstyleone}%
\theoremstyle{thmstyletwo}%

\theoremstyle{thmstylethree}%

\begin{document}

\title[Article Title]{Understanding is Compression}


\author[2,5,1]{\fnm{Ziguang} \sur{Li}}
\equalcont{These authors contributed equally to this work.}

\author[1,6]{\fnm{Chao} \sur{Huang}}
\equalcont{These authors contributed equally to this work.}

\author[5,1]{\fnm{Xuliang} \sur{Wang}}
\equalcont{These authors contributed equally to this work.}

\author[1,6]{\fnm{Haibo} \sur{Hu}}
\equalcont{These authors contributed equally to this work.}

\author[3]{\fnm{Cole} \sur{Wyeth}}

\author[1,5]{\fnm{Dongbo} \sur{Bu}}

\author[2]{\fnm{Quan} \sur{Yu}}

\author[2]{\fnm{Wen} \sur{Gao}}

\author*[4,5]{\fnm{Xingwu} \sur{Liu}}\email{liuxingwu@dlut.edu.cn}

\author*[3,5]{\fnm{Ming} \sur{Li}}\email{mli@uwaterloo.ca}

\affil[1]{\orgdiv{Institute of Computing Technology}, \orgname{Chinese Academy of Science}, \orgaddress{\state{Beijing}, \country{China}}}

\affil[2]{\orgname{Peng Cheng Laboratory}, \orgaddress{\city{Shenzhen}, \country{China}}}

\affil[3]{\orgdiv{School of Computer Science}, \orgname{University of Waterloo}, \orgaddress{\street{N2L 3G1}, \city{Waterloo}, \state{Ontario}, \country{Canada}}}

\affil[4]{\orgdiv{School of Mathematical Sciences}, \orgname{Dalian University of Technology}, \orgaddress{\city{Dalian}, \country{China}}}

\affil[5]{\orgname{Central China Institute of Artificial Intelligence}, \orgaddress{\city{Zhengzhou}, \country{China}}}

\affil[6]{\orgname{Ningbo Institute of Artificial Intelligence Industry}, \orgaddress{Ningbo}, \country{China}}

\abstract{
Modern data compression methods are slowly reaching their limits after 80 years of research, millions of papers,
and wide range of applications. Yet, the extravagant 6G communication speed requirement raises a major
open question for revolutionary new ideas of data compression.

We have previously shown all understanding or learning are compression, under reasonable
assumptions. Large language models (LLMs) understand data better than ever before. Can they help us to
compress data?

The LLMs may be seen to approximate the uncomputable Solomonoff induction. 
Therefore, under this new uncomputable paradigm, we present LMCompress. LMCompress
shatters all previous lossless compression algorithms, doubling the lossless compression ratios of 
JPEG-XL for images, FLAC for audios, and H.264 for videos, and quadrupling the compression ratio
of bz2 for texts. The better a large model understands the data, the better LMCompress compresses.
}




\maketitle
%\linenumbers
\section{Introduction}\label{sec1}

Before the reader starts to read this article, we invite you to reminisce about the everyday experiences: When you see a tiger, didn't you keep it in mind as a ``large cat"? When you see 3.141592 ..., didn't you just note it down as a $\pi$? When you see a bird, didn't you only focus on its unique features like size and color? 

Yes, you have compressed the data! What makes you achieve this is not an ingeniously designed algorithm, but your understanding. Understanding makes compression, which is the motivation of this paper. 

Data compression, either lossless or lossy, is an essential technology that underpins modern communication. Particularly, lossless compression allows the data to be perfectly reconstructed, which is indispensable for executable programs, text documents, genomics, cryptography, and multimedia archiving or production.

Numerous lossless compression methods have been developed, for example, ZIP, FLAC, PNG, and lossless H.264/H.265. These methods are largely confined to the information-theoretic framework established by C. Shannon over 80 years ago \citep{6773024shannon}, relying on various frequency-based considerations or other computable properties (see Supplementary material for more details). These compression approaches, although being 
 computationally tractable, have reached their limits after 80 years of research. 

% Data compression is an extremely important technology that underlies modern communication. Basically, data compression is either lossy or lossless, where lossless means that the compressed data can be exactly reconstructed. Lossless compression is indispensable in cases where no information loss is allowed or deviations from the original data would be unfavourable. Examples include executable programs, text documents, genomics, cryptography, multimedia archiving or production, and so on. Hence, this paper will address lossless compression of common types of data: text, audio, image, and video.

% Numerous methods for lossless ompression have been proposed and substantially transformed our life in the networked society. For example, ZIP and GZIP for general purpose, PPM for text, FLAC for audio, PNG and GIF for images, H.264/H.265 lossless and FFV1 for video. However, traditional compression has largely been confined to the information-theoretical  framework established by Shannon over 80 years ago \citep{6773024shannon}. They rely on various frequency-based concerns, Shannon entropy, or other computable properties. While being computable, such methods have reached their limits after 80 years of research.

Dawn of profound transformation appears with the advent of large models.
%have shown that compression and intelligence are two sides of the same coin \citep{deletang2023language}.
%, which has brought new hope for compression technology. 
The principle of large models dates back to the well-known Solomonoff induction proposed in 1960's \cite{rj1964formal}. Rather than extracting computable features, large models approximate the uncomputable Solomonoff induction from a lot of data. They understand the data in this way, hence enabling efficient compression as we do in everyday experience.

\begin{figure*}[!htb]
    \centering
    \includegraphics[width=\textwidth]{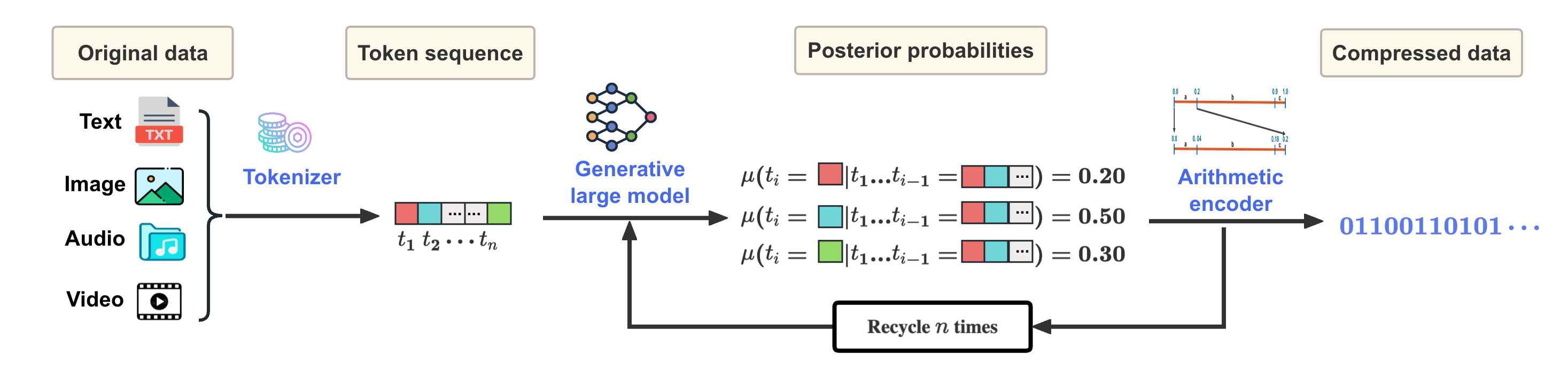}
    \caption{The architecture of our LMCompress. First, the original data is transformed into a sequence of tokens.
Then, this token sequence is fed into a generative large  model,
which outputs the predictive distribution for each token. Finally,
arithmetic coding losslessly compresses the original data
based on the predictive distributions. The
tokenization module and the generative large model may vary
according to the type of the data.}
    \label{fig:LMCompress}
\end{figure*}

% We will demonstrate significant lossless compression improvements on all types of data: 
% text, images, audios, and videos, 
% %than all previous models, several folds better than 
% %traditional algorithms, and a large margin better than other large model based compression.
% compared with all previous models. Our method is several folds better than traditional algorithms, and a large margin better than other large model based compression.

Based on the above observation, we advocate a new paradigm of compression, using large models to understand and consequently compress various data (see Fig. \ref{fig:LMCompress} for an overview).   
A precursor of our work has been independently published by us \cite{huang2023approximating} and by a DeepMind team \citep{deletang2023language}, with preceding work in~\cite{NNCP2021} with similar ideas. These works demonstrated that arithmetic coding with a generative large model can improve the best traditional text compressors such as gzip a few folds. %\cite{deletang2023language} also achieved the  state of the art of classical image compression, and moderately improved classical audio compression.

%\liu{Should we mention the multimedia compression in \cite{deletang2023language}?}

This paper intends to comprehensively justify the idea that better understanding implies better compression with a focus on lossless compression for clean comparisons.
To better understand the data with various formats, 
%demonstrate that better understanding implies better compression,
we use image GPT rather than a plain large language model (LLM) for images and videos, retrain an LLM with a small amount of audio data for audios, and employ domain-specific finetuned LLMs for domain texts. Based on the acquired understanding of data, we next apply arithmetic coding (see Supplementary material for details) to compress them. Lossless compression experiments show that by using LLM to understand data, we significantly improve compression ratios on all types of data, including texts, images, videos, and audios. Our method is several folds better than traditional algorithms, and a large margin better than the plain LLMs otherwise.
%used in our previous work in \cite{huang2023approximating} and in \cite{deletang2023language}.

%get significantly better compression ratios than existing methods on all types of data: texts, images, and audios, demonstrating the power of understanding on compression.
%We demonstrate the power of this idea by
% providing the significantly better 
% lossless compression ratios than previous best methods for 
% all types of data: texts, images, videos, and audios.
%We demonstrate that, according to our principles, the better the understanding, the better the compression, on images and audios. We thus used image GPT to significantly outperform the plain LLM on images, and we retrained an LLM with a small amount of audio data to greatly improve the compression on audio files. 

\section{Methods}\label{sec3}
Traditional compression methods, whether lossy 
or lossless, depend on a {\it computable}  function for characterizing data.  
Here, we propose LMCompress, a new Kolmogorov paradigm 
of compression rooted at the {\it uncomputable} Solomonoff 
induction. The Solomonoff induction is 
approximated by large models with never ending input data.
The compression ratio should go up with better approximation
of Solomonoff induction and better understanding of data.

It turns out that we have already passed the critical 
point as data accumulate and models improve. We demonstrate that LMCompress improves lossless compression 
of texts, images, videos and audios by several folds, far beyond the traditional methods.

The process of LMCompress is illustrated in Fig. \ref{fig:LMCompress}. 
First, we decompose the original data into a sequence of tokens. 
Then, we feed this token sequence into a large generative model, which outputs the predictive distribution for each token. 
Finally, we use arithmetic coding to losslessly compress the original data based on these predictive distributions. To better understand the data of various formats, we use different tokenization methods and large generative models for different data types, which are described in more detail as follows.

\subsection{Image Compression}\label{sec:imagecom}
% 先讲清楚大模型在图像语料上训练使它能够理解图像。通过自回归生成的方式，理解了图像表达的含义。这种理解型的压缩要比传统的压缩方案要好。传统的压缩方案利用统计规律进行压缩。
% For lossless compression, we need probability of each symbol to perform entropy coding algorithms such as arithmetic coding. Large models are the best  estimators of probability distributions. Hence,
% we consider an image as a sequence of pixels and model the image pixel distribution using a large vision 
% %language 
% model. Each pixel generates a token akin to text. Therefore, the probability model of image pixels can be the same as  a large language model, using the probabilities to generate the image pixel by pixel in an auto-regressive manner. Since large language models can learn semantic information  in text through auto-regression, similar image semantic information can also be learned in the vision language model. Then it can predict the probability of image pixels, and then compress the image through entropy coding. \par
%Consider an image $x \in \mathbb{R}^{H \times W \times 3}$ to be compressed. 

We use the image-GPT model (iGPT, \cite{chen2020igpt}) as the generative large model for images. Our choice of iGPT is driven by two key factors. 

First, iGPT is a large-scale vision model that has been trained on a vast corpus of images, equipped with a thorough understanding of visual data. This makes iGPT well-suited for analyzing and processing images.

Second, iGPT is an autoregressive large vision model. When presented with a sequence of pixels, it can generate predictive probability for each pixel in the sequence. This capability is a prerequisite for arithmetic coding.

To compress an image using iGPT, we first concatenate the image's pixels from top row to bottom row, transforming the two-dimensional visual data into a one-dimensional sequence of pixels. This pixel sequence is then fed into iGPT, yielding next-pixel probability for each pixel, based on which the image is compressed by using arithmetic coding.

The entire pixel sequence, however, cannot be fed into the model all at once due to the limited context window of iGPT model. In this case, we divide the sequence into non-overlapping segments, each of which can fit within iGPT's context window. These individual segments are then fed into iGPT and compressed independently.

\subsection{Video Compression}
\subsubsection{Lossless Video Compression}
% To the best of our knowledge, all existing open-source large video models are not autoregressive in nature.我搜索了一下自回归的视频视频大模型是有的，但是多数模型不能直接输出概率而是直接生成图像和视频，所以稍微改了一下。
To the best of our knowledge, all existing open-source large video models do not naturally output probabilities. To circumvent this limitation, we have opted  to leverage the image-based generative model iGPT instead. Since a video is fundamentally a sequence of frames, we propose to regard each frame as an image and compress the video frame-by-frame using iGPT as in Section \ref{sec:imagecom}.

At this stage, we have chosen not to exploit the inter-frame information for compression due to the following two concerns: 

First, many types of videos, such as action movies, exhibit drastic changes from one frame to the next. In such cases, attempting to leverage information from previous frames is unlikely to be effective for compressing the current frame.

Second, even for the videos with relatively modest inter-frame variations, such as classroom lecture recordings, we have found that utilizing the inter-frame information does not actually improve the overall compression performance. This may be because the iGPT model is already able to sufficiently understand and model each individual frame on its own.

By compressing each video frame independently using iGPT, we can sidestep the challenge posed by the lack of large autoregressive video models. This frame-by-frame compression approach allows us to harness the powerful image understanding capabilities of iGPT, thus exempting the need to address the complexities of modeling temporal dependencies between video frames.

\subsubsection{Lossy Video Compression}
Lossy compression is the main stream in video compression as loss is acceptable or even unavoidable in most scenarios of video application.  Numerous techniques have been proposed in this line, say DCVC series \citep{li2021deep} and H.26X series. Basically, all the traditional methods compress videos by removing intra- and inter-frame redundancies. Recently, the development of Artificial Intelligence Generated Content has inspired a new direction, namely, not only removing redundancies to compress, but also generating details to help reconstruction. This is pioneered by ``generative compression" \citep{santurkar2018generative}, and has attracted much attention in image compression\citep{yang2024lossy,relic2024lossy}. 

In this study, we extend the ``generative compression" idea to handle videos with an inspiration from \cite{xu2024idempotence}, which proposed a transform-coding based method linked with generative large model. Specifically, we use DCVC results as prior and sample from the diffusion model DDPM. The sampling process is modeled as an inverse problem by subtracting the squared error between each frame generated by the diffussion model and the DCVC-decoded frame in each iteration, similar to \cite{chung2022diffusion}. Because gradient of DCVC is necessary, we use the proxy loss function rather than the quantization step with additive uniform noise \cite{balle2016end}. More details are provided in Supplementary material.

\subsection{Audio Compression}
% Audios, as a type of sequential data, should undoubtedly exploit the sequential modeling ability of auto-regressive models. To capture long term patterns, state-of-the-art large audio models tend to discretize an audio into tokens, which is inevitably a lossy transformation. To achieve lossless compression, we establish a model that handles 
% audio at the level of signal, rather than tokens.
% \par
% Basically, the audio consists of a sequence of frames, each of which can be represented by a constant number of bytes. Note that a frame contains the amplitude information at a time point in the audio signal. We map each byte for a frame into an ASCII character, transforming the audio into a string of characters. What’s in need is just an auto-regressive model which understands this audio string. We implement such a model by adding a low rank adaption layer to an LLM and fine-tuning the LLM with audio strings. The fine-tuned model can estimate next-token probabilities for an audio string and thus compress the string with arithmetic coding.

Since audio is a type of sequential media, it is natural to leverage the large autoregressive models in compression. However, state-of-the-art open-source large audio models tend to discretize audio, which is inevitably lossy. To achieve lossless compression, we propose a model that handles audio at the signal level without discretization.

The basic idea is to treat audio as a sequence of frames. Each frame contains the amplitude information at a specific time point and can be represented by a constant number of bytes. We then map each byte into an ASCII character, effectively transforming the audio into a string of characters.

What we need is an autoregressive model that can understand this audio-as-string representation. To this end, we implement such a model by adding a low-rank adaptation layer to a large language model and subsequently fine-tuning the model on the acquired audio-as-string data. In this way, the fine-tuned model is able to estimate next-token probabilities for the audio string, which enables us to compress the audio using arithmetic coding.

Note that when the LLM has a limited context window, the audio string will be compressed piece-by-piece, as in Section \ref{sec:imagecom}. 

\subsection{Text Compression}
%It's relatively straightforward to come up with that LLMs may demonstrate even better compression ratio if the texts to be compressed are restricted in certain domains. By adding adaption layer, we enhanced the LLMs' ability to better understand and compress texts within certain domains. Given a piece of text with specified domain, we tokenize it and feed it into the domain-aware LLM. The LLM then generates the probability which will be used in arithmetic coding.

Large language models  have demonstrated impressive capability in compressing general texts \cite{huang2023approximating,deletang2023language}. Intriguingly, they have potential to achieve even better compression ratios, provided that the texts to be compressed are restricted to specific domains.

The key lies in adapting LLM to better understand the target domain. This is implemented by incorporating an adaptation layer and fine-tuning the LLM via domain-specific texts, tailoring the model to the characteristics of the domain.
%, resulting in a domain-aware LLM.

Then, to compress a text in a specific domain, we feed the text into the fine-tuned LLM. The LLM will estimate the next-token probabilities for the text, which can be leveraged by arithmetic coding to perform domain-specific compression. Again, when the LLM has a limited context window, the text will be compressed piece-by-piece, as in Section \ref{sec:imagecom}.

\section{Results}\label{sec4}

We use compression ratio as the metric of compression performance, which refers to the ratio of the size of the original data to that of the compressed data. In general, the bigger the compression ratio, the better the compression performance.

Most of the baselines, say H.264 video compression standard, can work in both lossy and lossless modes. For fair comparison, all the baselines are set to their lossless modes in our experiments, except for DCVC in lossy video compression.

In addition, the compression ratio of an algorithm varies across datasets, so every lossless compression algorithm is evaluated on two different datasets in order to mitigate the effect brought about by potential data biases. 

\subsection{Image Compression}
\textbf{Dataset} \quad We evaluate the image compression performance of LMCompress  on two benchmark image datasets, including $i)$ ILSVRC2017 \citep{russakovsky2015imagenet}, a large-scale dataset containing millions of labeled images across thousands of categories, derived from the ImageNet corpus, and $ii)$ CLIC2019 professional \citep{clicdataset}, which is specifically designed for evaluating image compression algorithms. CLIC2019 contains high-quality images with diverse characteristics such as natural scenes, textures, patterns, and structures, representative of real-world photography, multimedia, and visual content scenarios.

Since the datasets are too large, we sample 197 images from them. The total size of the images is 128 megabytes. Each image has three channels, corresponding to red, green, and blue, respectively, which are compressed separately. 
For each channel, we concatenate the rows of the image into a sequence, every 1024-pixel segment of which is fed into iGPT. Here the number 1024 is chosen to fit the context window of iGPT.
% baseline

Table \ref{tab:image1} 
%and Fig. \ref{fig:ICR} 
shows the compression ratios of the baselines and LMCompress on the two datasets. The results demonstrate that our LMCompress significantly outperforms all the baselines on both datasets, more than doubling the compression ratios. Note that Chinchilla is also a family of large models. LMCompress shows better performance than Chinchilla, possibly because Chinchilla is only trained on language corpus while LMCompress is trained on image corpus hence can understand images better.

%JPEG-2000 \citep{jpeg2000}.\par
\begin{table}[h]
\caption{Image compression ratios of state-of-the-art compressors and our LMCompress. Datasets: CLIC2019 and ISLVRC}
\begin{tabular}{@{}lll@{}}
\toprule
Method & CLIC2019 & ISLVRC \\
\midrule
PNG & 2.205\footnotemark[1] & 1.67   \\
JPEG-XL & 2.93\footnotemark[1] & 1.90  \\
WebP & 2.75\footnotemark[1] & 2.04  \\
JPEG-2000 & 2.73\footnotemark[1] & 1.53 \\
Chinchilla 7B & $\backslash$ & 1.82\footnotemark[2] \\
Chinchilla 70B & $\backslash$ & 2.08\footnotemark[2] \\
LMCompress & 6.32 & 4.79  \\ 
\botrule
\end{tabular}
\footnotetext[1]{These results are from \citep{rhee2022lc}}
\footnotetext[2]{These results are from \citep{deletang2023language}. Results on CLIC2019 are not presented since neither
such results nor Chinchilla models are publicly available.}
\label{tab:image1} 
\end{table}

\begin{figure}
    \centering
    \includegraphics[width=0.80\linewidth]{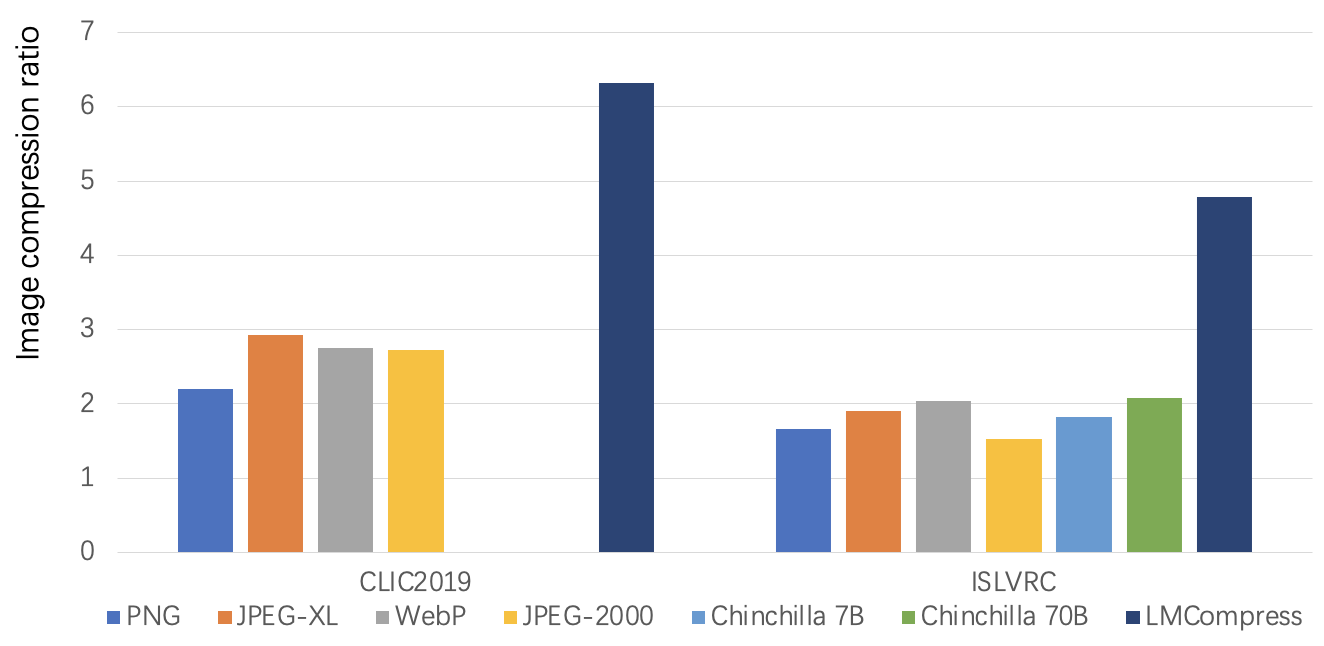}
    \caption{Image compression ratios}
    \label{fig:ICR}
\end{figure}

% The parameters of the large models can impact their performance. Therefore, we evaluate the effectiveness of LMCompress by testing different windows and data extraction methods on the CLIC dataset for a more comprehensive evaluation of our method. The results are shown in Table \ref{tab:image2}. In the table, ``patch" means that the image will be divided into different patches which will be sequentially fed into the model for computation, while ``sequence" means that the image will be turned into a one-dimensional sequence of pixels. We observe extracting data in the form of sequences achieves better compression ratios than in the form of patches. The reason is that the sequence method is more in line with the way of understanding images during model pre-training. This phenomenon is consistent with our point of view that understanding is compression. Better understanding of data implies better compression ratios.

% \begin{table}[h]
%     \centering
%     \begin{tabular}{c|c|c}
%         \hline
%         % extraction method context window size
%         windows size & CLIC \\
%         \hline
%         LMCompress-1024  & 6.55 \\ 
%         LMCompress-512  & 5.33\\
%          \hline
%     \end{tabular}
%     \caption{Compression ratios using different window sizes.}
%     \label{tab:image2}
% \end{table}

\subsection{Video Compression}
\subsubsection{Lossless Video Compression}
\textbf{Datasets} We use test data from  Xiph.org, which has over 1000 video clips in the uncompressed YUV4MPEG format. Since the whole dataset is too massive, we sample 10 video clips as test data: 5 of static scenes and 5 of dynamic scenes. A static scene means that the consecutive frames change slightly or not change at all, for example, classroom recordings. The total size of static-scene videos is 162 megabytes. On the contrary, a dynamic scene means that the frames change drastically, for example, motion movies. The total size of dynamic-scene videos is 237 megabytes. 

%\begin{table}[h]
%\caption{Lossless video compression ratios}
%\begin{tabular}{@{}lll@{}}
%\toprule
%Method & static scene &  dynamic scene \\
%\midrule
%FFV1  & 2.63 & 2.13 \\ 
%H.264  & 3.38  & 2.53\\
%H.265  & 4.75  & 3.51 \\ 
%LMCompress  & 5.60  & 5.59 \\ 
%\botrule
%\end{tabular}
%\label{tab:lossless_video} 
%\end{table}

\begin{figure}
    \centering
    \includegraphics[width=0.6\linewidth]{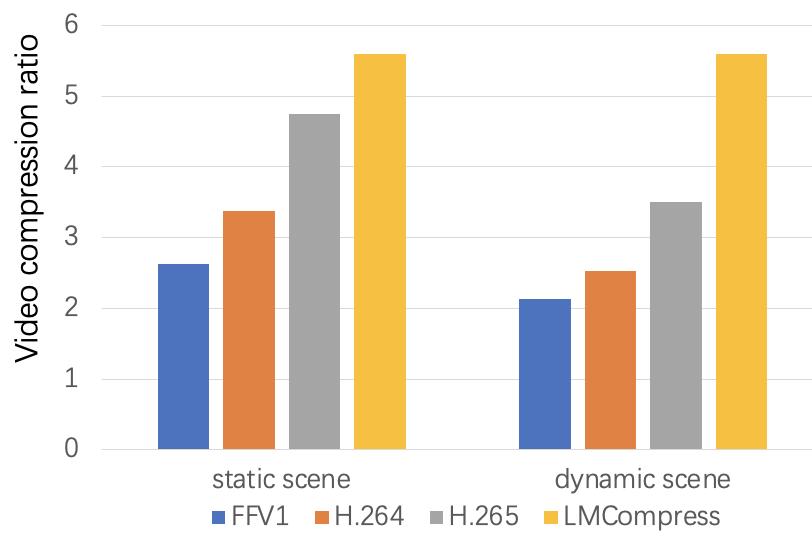}
    \caption{Lossless video compression ratios of the state-of-the-art approaches and our LMCompress. Dataset: Xiph.org videos classified into ``static scene" and ``dynamic scene" }
    \label{fig:VCR}
\end{figure}

\par
As shown in %Table~\ref{tab:lossless_video} and 
Fig. \ref{fig:VCR}, LMCompress outperforms both baselines on both video types. 
On static scenes, LMCompress achieves over 20\% improvement in compression ratio compared to the baselines. Even on dynamic scenes,  LMCompress maintains its edge, achieving at least 50\%  improvement in compression ratio.
We further observe that dynamic scenes are harder to compress than static scenes. A possible reason is that in a dynamic-scene video, the actors tend to be in transient postures which are hard to understand or predict.

\subsubsection{Lossy Video Compression}
\textbf{Datasets} \quad 
% We use CIPR SIF Sequences at Xiph.org to evaluate LMCompress. Due to the resolution limitation of the diffusion model, we scale the video size to $256\times 256$.
%
% \textbf{Metric} \quad In adition to the main metric compression ratio, we also adopt three other metrics widely used to measure video quality: Peak-Signal-Noise-Ratio (PSNR) to measure distortion, bpp (bits per pixel) to measure bitrate and Fr\'echet Inception Distance (FID) to measure perceptual quality.
%
% We follow the diffusion training setting and hyperparameters in \citep{chung2022diffusion}, which uses stochastic gradient descent to optimize the intermediate samples. The forward measurement operator is as in DCVC\citep{li2021deep}. We choose ELIC\citep{he2022elic} as I-frame compressor.
We utilize CIPR SIF Sequences from Xiph.org as our evaluation dataset for LMCompress. However, due to the limitations of the diffusion model's resolution, we rescale the video size to $256\times 256$.

\textbf{Metrics} In addition to the primary metric of compression ratio, we also adopt three widely used metrics to assess video compression quality. These metrics include Peak-Signal-Noise-Ratio (PSNR) for measuring distortion, bits per pixel (bpp) for bitrate, and Fr\'echet Inception Distance (FID) for perceptual quality.

We follow the diffusion training setting and hyperparameters outlined in \citep{chung2022diffusion}, which employs stochastic gradient descent for optimizing intermediate samples. The forward measurement operator utilized is the same as in DCVC \citep{li2021deep}. We choose ELIC \citep{he2022elic} as the I-frame compressor.

\begin{table}[h]
\caption{Lossy video compression performance of the state-of-the-art approaches and our LMCompress. Dataset: CIPR SIF Sequences}
\begin{tabular}{@{}lllll@{}}
\toprule
&Compression ratio & bpp & PSNR $\uparrow$ & FID $\downarrow$ \\
\midrule
DCVC & 162 & 0.0945 & 29.0 & 153\\
DCVC-FM & 269 & 0.0569 & 31.8 & \ \\
LMCompress & 582 & 0.0263 & 32.3 & 81\\
\botrule
\end{tabular}
\label{tab:lossyVideo}
\end{table}

%\begin{figure}
%    \centering
%    \includegraphics[width=0.75\linewidth]{figures/VCRL.png}
%    \caption{Lossy vedio compression performance}
%    \label{fig:VCRL}
%\end{figure}

As demonstrated in Table \ref{tab:lossyVideo}, 
%and Fig \ref{fig:VCRL}, 
LMCompress exhibits superior performance compared to DCVC and DCVC-FM. Notably, LMCompress more than doubles the compression ratio, while keeping the other metrics as good or even better.

\subsection{Audio Compression}
\textbf{Dataset} \quad We use LibriSpeech ASR corpus\citep{Vassil2015Librispeech} and LJSpeech \citep{ljspeech17} as datasets to test audio compression. Both datasets are collected from the LibriVox project which covers nearly 1000 hours of 16 kHz English speech in audiobooks. Since the datasets are too large, we extract the first Gigabytes from the \emph{train-clean-100} split of the LibriSpeech corpus and the first 256 Megabytes from LJSpeech. 

To be processed by the LLMs in our experiment, each audio clip is transformed into a string of ASCII characters. Specifically, we right shift one bit for every byte in the frames so that each byte is a valid ASCII character. The discarded bits due to the shifts are stored independently. The strings are then divided into 2048-byte chunks, each of which is fed into the LLMs and compressed separately. Here, the number 2048 is chosen to fit in the context window of the LLMs.

%The audio streams are transformed into strings of ASCII characters before being compressed. The strings are further divided into non-overlapping segments of 2048 bytes, so that every segment does not exceed the context window size of the LLaMA models. The segments are compressed independently.\par
\textbf{Model fine-tuning} \quad We build LMCompress for audio compression by conducting supervised LoRA\citep{hu2022lora} fine-tuning on the LLaMA series model\citep{Hugo2023LLaMA} LLaMA3-8B. Note that LLaMA3-8B was pretrained on normal texts. To tailor it for audio compression, we fine-tune it on a corpus consisting of ASCII character strings derived from audio clips. For this end, we choose the first 64 megabytes in the \emph{dev-clean} split of LibriSpeech as training data, with rank 8 and alpha 32.  

\begin{table}[h]
\caption{Audio compression ratios. Dataset: LibriSpeech and LJSpeech}
\begin{tabular}{@{}lll@{}}
\toprule
Method & LibriSpeech & LJSpeech \\
\midrule
FLAC & 3.23 & 3.21\\
LLaMA3-8B & 4.45 & 4.02\\
Chinchilla-7B & 4.24\footnotemark[1] & \textbackslash \\
Chinchilla-70B & 4.76\footnotemark[1] & \textbackslash \\
LMCompress & 6.07 & 6.22 \\
\botrule
\end{tabular}
%\footnotetext[1]{\cite{deletang2023language} experimented compression with LLaMA2-7B model. Here we implemented the same method on top of LLaMA3-8B model, since our proposed LMCompress is trained from LLaMA3-8B model.}
\footnotetext[1]{These results are from \cite{deletang2023language}. Results on
LJSpeech are not presented since neither
such results nor Chinchilla models are publicly available.}
\label{tab:audio}
\end{table}

%\begin{figure}
%    \centering
%    \includegraphics[width=0.75\linewidth]{figures/ACR.png}
%    \caption{Audio compression ratios}
%    \label{fig:ACR}
%\end{figure}

The results are shown in Table \ref{tab:audio}. %and Fig. \ref{fig:ACR}. 
We obtained two observations: $i)$ Large-model based methods outperform the state-of-the-art traditional method FLAC by 25\%-94\%, and $ii)$ LMCompress, which was fine-tuned on audio corpus, outperforms other large-model based methods, with margins from 28\% to 55\%. Surprisingly, even though fine-tuned on the dataset LibriSpeech only, LMCompress  improves the compression ratio of the raw  LLaMA3-8B on LJSpeech by 55\%.

% According to Table \ref{tab:audio}, even fine-tuned with a small amount of audio data, the model generalizes well on much larger test datasets. LMCompress almost doubles the compression ratio of the classic FLAC method. Furthermore, it outperforms raw LLaMA3-8B by 36\% on LibriSpeech and by 55\% on LJSpeech, respectively. The results support our observation that better understanding leads to better compression.

% Note that no data from the LJSpeech dataset is involved in the fine-tuning process.

\subsection{Text Compression}
\textbf{Dataset} \quad Our benchmarks for domain-aware text compression are the MeDAL\citep{wen2020MeDAL} and the Pile of Law\citep{hendersonkrass2022pileoflaw}. 
MeDAL is a dataset in the domain of medicine. It is created from PubMed abstracts which are released in the 2019 annual baseline and primarily serves as a corpus for medical abbreviation understanding.
On the other hand, Pile of Law is a dataset in the domain of law. It includes legal and administrative texts compiled from 35 sources.

%The domains of the datasets are medicine and law, respectively. Specifically, MeDAL is created from PubMed abstracts which are released in the 2019 annual baseline and primarily serves as a corpus for medical abbreviation understanding. Pile of Law is a dataset of legal and administrative texts compiled from 35 sources. 

In the experiments, we extract the first 1104 Megabytes from MeDAL and the \emph{eurlex} split from the Pile of Law corpus. Again, we divide the texts into segments of 2048 bytes so that every segment fits the context window of the large language models. 

%The segments are compressed independently.\par
\textbf{Model fine-tuning} \quad We build LMCompress for domain-aware text
compression by fine-tuning LLaMA3-8B via supervised LoRA. 
For either domain dataset, we use the first 64 Megabytes for training, the next 16 Megabytes for validation, and the rest for testing. 

% \begin{table}[h]
% \caption{Text compression ratios}
% \begin{tabular}{@{}lll@{}}
% \toprule
% Method & MeDAL & Pile of Law \\
% \midrule
% zlib & 2.96 &  3.14\\
% bzip2 & 3.94 & 4.15\\
% brotli\citep{Jy2019Brotli} &  4.22 & 5.13 \\
% LLaMA3-8B\footnotemark[1] & 9.66 & 12.15 \\
% LMCompress & 10.48 & 16.81 \\
% \botrule
% \end{tabular}
% \footnotetext[1]{This is adapted from the text compressor in \citep{huang2023approximating} by replacing LLaMA2-7B with LLaMA3-8B.}
% \label{tab:text}
% \end{table}

\begin{figure}
    \centering
    \includegraphics[width=0.75\linewidth]{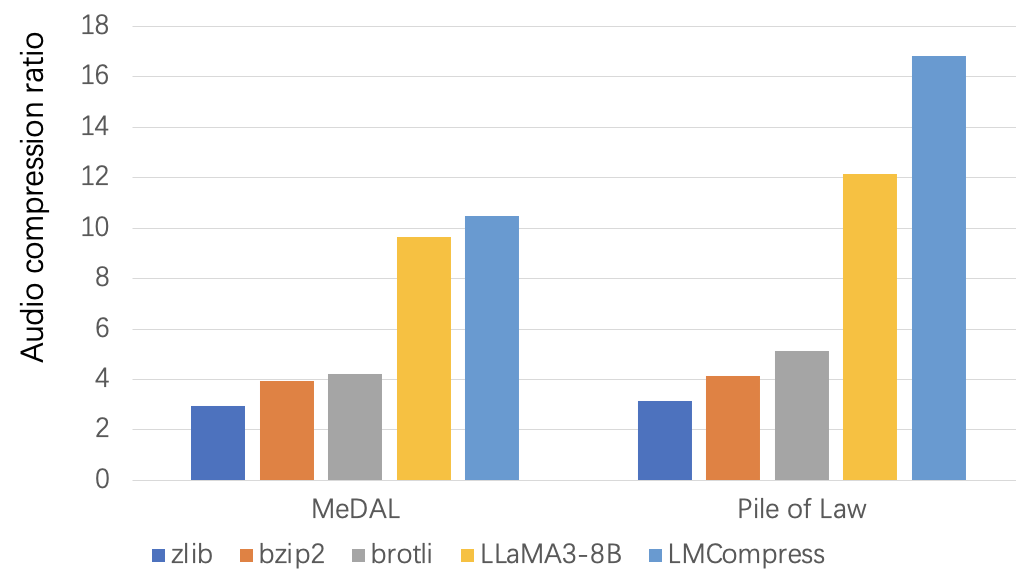}
    \caption{Text compression ratios. Dataset: MeDAL and Pile of Law. LLaMA3-8B means the text compressor in \citep{huang2023approximating} with LLaMA2-7B replaced by LLaMA3-8B}
    \label{fig:TCR}
\end{figure}

%The results are shown in %Table \ref{tab:text} and 
Fig. \ref{fig:TCR} suggests that LMCompress outperforms all the baseline approaches. Its compression ratio on either dataset almost triples those of the best traditional methods. Compared to raw LLaMA3-8B, LMCompress improves the compression ratio by 8.5\% on MeDAL and by 38.4\% on Pile of Law.

%Table 4 not only gives a solid guarantee for LLMs' compression performance on texts, but also proves that better understanding of the texts' domain helps to compress.

%\textbf{To sum}, 
In summary, LMCompress has higher lossless compression ratios on various media than all traditional baselines and raw LLM-based algorithms. This evidence supports our claim that better understanding leads to better compression.

\section{Conclusion}\label{sec5}
Communication in the past was generally governed by the Shannon paradigm, 
with coding efficiency upper bound by Shannon entropy. 
While exploring other computable features can further improve
compression, large models may be seen to approximate the 
uncomputable Solomonoff induction, hence 
opening a new Kolmogorov paradigm of compression. 
As we have shown, this new way of lossless compression
has achieved substantial improvements on various kinds of data. 
This new paradigm allows us to systematically understand 
the data we transmit, shattering the Shannon entropy upper bound in a great scale. 

%Communication in the past was generally governed by the Shannon paradigm, 
% with compression ratio upper bound determined by Shannon entropy. 
% While exploring other computable features can further improve
% compression, large models may be seen to approximate the 
% uncomputable Solomonoff distribution hence 
% opening a new Kolmogorov paradigm of compression. 
% As we have shown, this new way of lossless compression
% has achieved several folds of improvements on various kinds of data. 
% This new Kolmogorov paradigm allows us to systematically understand 
% the data we transmit, shattering the Shannon entropy upper bound in a great scale. 

Our work sheds light on the 6G communication, especially when 
the bandwidth is limited from the satellites. It will be significantly benefited by understanding the data, with large models at both ends of communication to encode and decode.
As the large models are specialized as agents, assisted with Retrieval-Augmented Generation, AI will understand
the data to be transmitted much better. 
When the data need to be encrypted, our compression needs to be done before
encryption. One can even imagine that the sides with
superior models broadcast openly compressed messages, allowing only 
those with equal models to decipher as a first level of encryption, 
at no extra cost. 

Though this paper is mainly on lossless compression, our work on lossy video compression indicates that the research presented here
can be extended to the domain of lossy compression. Another future direction is to incorporate inter-frame information in lossless video compression. 

\backmatter

\bmhead{Acknowledgements}
This research is partially supported by Canada’s NSERC grant OGP0046506, and Canada Research Chair Program.
We thank Nick Zhang and Paul Vitanyi for discussions on Solomonoff
induction. We thank Cynthia Huang, Yuqing Xie, Zhiying Jiang,
Rui Wang, and Peijia Guo for their discussions and related work in 
\citep{jiang2023theory} and \citep{huang2023approximating}.

\bmhead{Data availability}
LSCRVC 2012 is available at \url{https://www.image-net.org/challenges/LSVRC/2012/index.php}. 
CLIC is available at \url{https://clic.compression.cc/2019/}. Librispeech is available at \url{www.openslr.org/12}. 
LJSpeech is available at \url{https://keithito.com/LJ-Speech-Dataset}. 
MeDAL is available at \url{https://github.com/McGill-NLP/medal}.  
Eurlex is available at \url{https://huggingface.co/datasets/pile-of-law/pile-of-law}. 
CIPR SIF Sequence (foreman) is available at \url{https://media.xiph.org/video/derf/}.

\bmhead{Code availability}
The code has been uploaded to Code Ocean, and will be publicly available at Github after the manuscript is accepted.

\bibliography{sn-bibliography}% common bib file

\end{document}